\documentclass[aps,showpacs,preprint,superscriptaddress]{revtex4}
\usepackage{graphics}
\usepackage{color}
\usepackage{epsfig}
\usepackage{times}

\newcommand{\cici}{{\rm\raisebox{-2.pt}{%
      $\stackrel{\scriptstyle\circ}{\scriptstyle\circ}$}}}

\begin{document}      
      
\title{Test of the quantumness of atom-atom correlations in a bosonic gas}

\author{D.~Ivanov}

\affiliation{Laser Research Institute, St.~Petersburg State University,
  5 Uliyanovskaya, Petrodvoretz, St.~Petersburg, Russia} 
  
\author{S.~Wallentowitz}
    
\affiliation{Facultad de F{\'i}sica, Pontificia Universidad Cat{\'o}lica de
  Chile, Casilla 306, Santiago 22, Chile}

\begin{abstract}
  It is shown how the quantumness of atom-atom correlations in a trapped
  bosonic gas can be made observable. Application of
  continuous feedback control of the center of mass of the atomic cloud is
  shown to generate oscillations of the spatial extension of the cloud, whose
  amplitude can be directly used as a characterization of atom-atom
  correlations. Feedback parameters can be chosen such that the violation of a
  Schwarz inequality for atom-atom  correlations can be tested at noise
  levels much higher than the standard quantum limit.
\end{abstract}


\date{30 September 2005}

\maketitle

In optics correlations between photons in a light field have been demonstrated
using a Hanbury Brown Twiss setup~\cite{hanbury-brown-twiss}. There the
incoming light beam is split by a semitransparent mirror and the two outputs are
measured with photodetectors. By displacing detectors normal to the detected
beams and by introducing a variable time delay, from the coincidences of
detections the intensity-intensity correlation of the quantized light field,
\begin{equation}
  \label{eq:c-intro}
  C(1,2) = \big\langle \cici \, \hat{I}(1) \, \hat{I}(2) \, \cici
  \big\rangle .
\end{equation}
is obtained, where $1$ and $2$ represent the space-time coordinates $(x_1,t_1)$
and $(x_2,t_2)$, respectively. Here $\hat{I} \!=\! \hat{E}^{(-)}
\hat{E}^{(+)}$ with $\hat{E}^{(\pm)}$ being positive- and negative-frequency
parts of the (here scalar) electric field of the light beam and $\cici \; \cici$
denotes time and normal-ordering.

Interpreting the expectation value in Eq.~(\ref{eq:c-intro}) as being based on a
proper probability density, application of the Cauchy--Schwarz inequality leads
to
\begin{equation}
  \label{eq:cs-intro}
  C(1,2) \leq \sqrt{ C(1,1) \, C(2,2)}.
\end{equation}
This inequality is based on the assumption of a classical random process
with correlations describing either complete randomness or bunching of
photons~\cite{mandel-book}. However, quantum states of light exist that cannot
be described in this way and violate the inequality~(\ref{eq:cs-intro}).
This has been used as a criterion for true quantumness --- or non-classicality
--- of the incoming light beam, and has been experimentally observed
as antibunched
light~\cite{antibunching1,antibunching2,antibunching3,antibunching4}.

These concepts can also be applied to ultracold bosonic gases,
where the electric field is replaced by the matter-field
$\hat{\phi}(x,t)$. However, different from optical correlations measured by
absorbing photodetectors, atom correlations will not occur in normal operator
ordering. Nevertheless, an analogy between photonic and atomic correlations can
be drawn. In fact, atom-atom correlations have been recently measured in a way
analogous to the Hanbury Brown--Twiss optical setup~\cite{yasuda,bloch,aspect}.
Furthermore, there have been also approaches where the measurement of losses due
to three-body recombinations has been used to infer on those
correlations~\cite{kogan,holland,phillips}.

In these techniques either the gas has to be released from the trap or atomic
losses play a major role. One might, however, manipulate or drive the system in
such a way as to map the internal atom-atom correlations into easily accessible
mesoscopic observables, that may be observed {\it in situ} without loosing atoms
from the trap. This is the approach taken in this Letter: By applying a
continuous feedback control of the atomic cloud's center of mass and by
observing the cloud size, it is shown that atom-atom correlations can be
detected and used to test for the quantumness of atom-atom correlations, in
close analogy to antibunching in the case of photons.

Consider an ideal bosonic gas with $N$ atoms of mass $m$, that is kept in a
harmonic trap of frequency $\omega$. A feedback loop is continuously applied to
compensate for the motion of the center of mass of the gas $\hat{X}$, i.e. to
damp the collective motion of atoms. It consists of the continuously repeated
application of a measurement of $\hat{X}$ with measurement outcome $X_{\rm m}$
and resolution $\sigma_0$, and a corresponding shift by $-\zeta_0 X_{\rm m}$.
Both processes shall be much faster than the free oscillation of the system with
trap frequency $\omega$, so that an instantaneous action can be assumed.

Experimental implementations of such a feedback loop are based on the collective
interaction of the atoms with far off-resonant optical probe fields.
Two-photon transitions, whose strength depend on the positions of atoms in
the probe field, then lead to a redistribution of intensities in optical-field
modes. The latter may be detected and used as an input signal for a subsequent
control action using optical phase shifts of the same probe field.
Such a scheme has been realized with atoms in optical lattices~\cite{raithel} or
may be realized as an extension of the single-atom experiment by Fischer {\it
et~al.}~\cite{rempe}. Possibly it may also be integrated into the
magneto-optical or optical trap configuration, by detecting and modulating the
trapping laser fields.

Given the continuous application of feedback at rate $\gamma$, two
parameters determine the feedback dynamics: the rms time-integrated measurement
resolution $\sigma \!=\! \sigma_0 / \sqrt{\gamma}$ and the feedback shift rate
$\zeta \!=\! \zeta_0 \gamma$. The time evolution of the $N$-atom density
operator of the system, $\hat{\varrho}_N$, is then described by the master
equation of quantum Brownian motion~\cite{master1,master2,master3},
\begin{equation}
  \label{eq:N-master}
  \partial_t \, \hat{\varrho}_N
  = - \frac{i}{\hbar} [ \hat{H}, \hat{\varrho}_N
  ]  + i \frac{\zeta}{2\hbar} [ \hat{P}, \{ \hat{X}, \hat{\varrho}_N
  \} ]
  - \frac{1}{8\sigma^2} [ \hat{X}, [\hat{X}, \hat{\varrho}_N]]
  - \frac{\zeta^2\sigma^2}{2\hbar^2} [ \hat{P}, [ \hat{P},
  \hat{\varrho}_N]] .
\end{equation}
Here $\hat{H}$ is the Hamiltonian of the non-interacting atomic gas in the trap
potential and the center-of-mass (cm) operator and its canonically conjugate
total momentum read
\begin{equation}
  \label{eq:XP-def}
  \hat{X} = \frac{1}{N} \int \! dx \, \hat{\phi}^\dagger(x) \, x \,
  \hat{\phi}(x) , \quad
  \hat{P} = -i\hbar \int \! dx \, \hat{\phi}^\dagger(x)
  \, \partial_x \, \hat{\phi}(x) ,
\end{equation}
with the atomic field $\hat{\phi}(x)$ obeying the bosonic commutator
relation $[\hat{\phi}(x), \hat{\phi}^\dagger(x')] \!=\! \delta(x \!-\! x')$.

The stationary behavior of the cm, obtained from
Eq.~(\ref{eq:N-master}), is an exponential damping at rate $\zeta/2$ of the
coherent oscillation: $\lim_{t\to\infty} \langle \hat{X}(t) \rangle \!=\! 0$.
Moreover, its rms spread converges exponentially at the same
rate to the non-zero stationary value
\begin{equation}
  \label{eq:rms-cm-limit}
  \lim_{t\to\infty} \sqrt{\langle [\Delta \hat{X}(t)]^2 \rangle} =
  \Delta X_{\rm s} ,
\end{equation}
where $\Delta \hat{X} \!=\! \hat{X} \!-\! \langle \hat{X} \rangle$.
It represents the noise left in the cm after a time of the order of $\zeta^{-1}$
needed for damping the coherent cm oscillation. This noise is determined solely
by the parameters of the feedback and trap:
\begin{equation}
  \label{eq:DX}
  \Delta X_{\rm s} = \delta X_0 \, \sqrt{(\eta \!+\!
  \eta^{-1}) / 2 } ,
\end{equation}
with the rms spread of the cm in the ground state of the trapping potential
being
\begin{equation}
  \label{eq:dX0}
  \delta X_0 = \sqrt{\hbar/(2 N m \omega)} ,
\end{equation}
which serves as the standard quantum limit (SQL) for the cm
coordinate. The parameter
\begin{equation}
  \label{eq:eta}
  \eta =  {\delta X_0}^2 / (\zeta\sigma^2)
\end{equation}
specifies the ratio of spatial localization due to the potential over that due
to the feedback. Note, that Eq.~(\ref{eq:DX}) attains the SQL as a minimum
value for $\eta \!=\! 1$ but is otherwise much larger.

Our goal is to describe atomic correlation effects in the dynamics of the
atomic density. For that purpose we need the single-atom density matrix
\begin{equation}
  \label{eq:density}
  \rho(x,x',t) =  \langle \hat{\phi}^\dagger(x') \, \hat{\phi}(x) \rangle_t ,
\end{equation}
with $\langle \ldots \rangle_t \!=\! {\rm Tr}[ \ldots \hat{\varrho}_N(t)]$.
To obtain the dynamical evolution of this density matrix from
Eq.~(\ref{eq:N-master}) of course is prevented by the correlations in
the many-atom systems. That correlations play a role in
Eq.~(\ref{eq:N-master}) can be seen from the occurrence of products of
operators $\hat{X}$ and $\hat{P}$, that contain products of four field
operators [cf.~Eq.~(\ref{eq:XP-def})], similar to atom-atom interactions.
However, recently it has been shown~\cite{ivanov-wal}, that despite of these
problems, the single-atom density matrix can be obtained via a procedure that we
may briefly outline here:

Instead of the single-atom density matrix, the joint Wigner function of single
atom (variables $x$, $p$) and cm of the other $N \!-\! 1$ atoms
(variables $X$, $P$) is considered:
\begin{eqnarray}
  \label{eq:wigner}
  W(x,p; X,P,t) & = & (2\pi\hbar)^{-3}
    \! \int \!  dx' \!  e^{-i x'p / \hbar}
    \int \! dX' \! \int \! dP' \nonumber \\
  & & \times \left\langle \hat{\phi}^\dagger \!\Big( x \!-\!
    \frac{x'}{2} \Big) \,
    e^{ i [ (\hat{P} - P) X' + (\hat{X} - X)  P'] / \hbar }
    \hat{\phi} \!\Big( x \!+\! \frac{x'}{2} \Big) \right\rangle_t . \qquad
\end{eqnarray}
From Eq.~(\ref{eq:N-master}) a closed Fokker-Planck equation follows for this
distribution, which is of linear type with positive semi-definite diffusion
matrix, leading thus to a bound analytic Green function of Gaussian
type~\cite{risken}. In consequence, given the initial conditions, analytic
solutions for this Wigner function can be obtained, from which the solution
for the single-atom density matrix are derived by integration over the
auxiliary phase-space variables:
\begin{equation}
  \label{eq:rho-W}
  \rho(x \!+\! x', x \!-\! x',t) = \int \! dX \! \int \! dP \!
  \int \! dp \, W(x,p;X,P,t) \, e^{2ip x'} .
\end{equation}

Thus in principle the complete atomic density profile could be obtained. Here
we focus on the rms spread of the corresponding atomic density,
\begin{equation}
  \Delta x(t) = \left\{ \int \! \frac{dx}{N} \, x^2 \, \rho(x,x,t) -
  \left[ \int \! \frac{dx}{N} \, x \, \rho(x,x,t) \right]^2
  \right\}^{\frac{1}{2}} ,
\end{equation}
giving us information on the quantum-statistically averaged temporal evolution
of the extension of the atomic cloud. From a complete solution of the
Fokker--Planck equation for (\ref{eq:wigner}) this variance can be shown to
exponentially converge at rate $\zeta/2$ to the asymptotic behavior
$\lim_{t\to\infty} \Delta x(t) \!\sim\! \Delta
x_{\rm a}(t)$, defined by
\begin{equation}
  \label{eq:dx}
  \Delta x_{\rm a}(t) = \sqrt{ {\Delta X_{\rm s}}^2 +
    {\sigma_q}^2(t) } .
\end{equation}

In this equation the first term in the square root is
given by the constant stationary rms cm spread [cf.~Eq.~(\ref{eq:DX})],
whereas the second term is explicitely time dependent and reads
\begin{equation}
  \label{eq:C-def}
  {\sigma_q}^2(t)
  = \int \! \frac{dx}{N} \left\langle \hat{\phi}^\dagger(x)
    \left[ q(x, t) \right]^2 \hat{\phi}(x) \right\rangle_0
  - \int \! \frac{dx}{N} \int \! \frac{dx'}{N} \left\langle
  \hat{\phi}^\dagger(x)
    \, q(x, t) \, \hat{\phi}(x) \, \hat{\phi}^\dagger(x') \,
    q(x',t) \, \hat{\phi}(x') \right\rangle_0 ,
\end{equation}
with the expectation value being taken with respect to the initial $N$-atom
density operator $\hat{\varrho}_N(0)$. The explicit time dependence of
Eq.~(\ref{eq:C-def}) is given by the single-atom quadrature, defined as
\begin{equation}
  \label{eq:q}
  q(x,t) = x \cos(\omega t) - \frac{i\hbar
  \partial_x}{m\omega} \sin(\omega t) .
\end{equation}
Equation~(\ref{eq:C-def}) represents the central result of our Letter. In its
second part it contains an atom-atom correlation function with four
matter-field operators, showing that due to the feedback these correlations
show up in the observable mesoscopic size of the cloud, cf.~Eq.~(\ref{eq:dx}).
Thus the atom-atom correlations will become visible in a purely single-atom
property.

Due to its explicit time dependence, in general there will be no
stationary size of the atomic cloud, but instead the cloud will
periodically breath. It should be emphasized, that this breathing has nothing in
common with the well-known collective oscillations of a Bose gas or condensate.
The latter rely on the presence of atomic collisions, whereas the
breathing discussed here is an effect solely produced by the feedback. Feedback
of course effectively mediates interactions between atoms, so that a certain
analogy to true atom-atom interactions can be drawn. However, whereas collisions
of ultracold atoms lead to Hamiltonian terms, the feedback in addition provides
non-unitary parts of the time evolution [cf.~Eq.~(\ref{eq:N-master})].

Thus, after a transient behavior during a time of the order of $1/\zeta$, the
cloud size will oscillate at twice the trap frequency, which resembles
single-atom quadrature squeezing~\cite{squeezing}. The amplitude of this
oscillation will of course depend on the initial quantum state and its
correlations at time $t\!=\!0$, before the feedback has been turned on.
Since the SQL for a single atom in the trap reads $\delta x_0 \!=\! \delta X_0
\sqrt{N}$, it follows that the atomic cloud size~(\ref{eq:dx}) can become
smaller than that value. The condition for quadrature squeezing (QS) on the
single-atom level would then be that at some time during the half period
$\pi/\omega$ the following inequality is fulfilled:
\begin{equation}
  \label{eq:squeeze-cond}
  \Delta x_{\rm a}(t) < \delta x_0 \qquad \mbox{(single-atom QS)}.
\end{equation}

However, there is more to Eq.~(\ref{eq:dx}) than single-atom QS.
It can be revealed by taking a closer look at the structure of
Eq.~(\ref{eq:C-def}) and applying the ideas developed in the context of photon
antibunching. The probability density for finding an atom at position $x$ is
undoubtedly defined as
\begin{equation}
  \label{eq:P(x)}
  P(x) = \langle \hat{\phi}^\dagger(x) \hat{\phi}(x) \rangle / N .
\end{equation}
A quasi joint probability density for two atoms being at positions $x$ and $x'$,
that is consistent with the definition~(\ref{eq:P(x)}), reads
\begin{equation}
  \label{eq:P(x,x')}
  P(x,x') = \langle \hat{\phi}^\dagger(x) \hat{\phi}(x)
  \hat{\phi}^\dagger(x') \hat{\phi}(x') \rangle  / N^2 .
\end{equation}
Note that different from the optical case this correlation is not normally
ordered. Consistency means here that for $N$ atoms the marginals of $P(x,x')$
reproduce the correct probability density:
\begin{equation}
  \int \! dx' \, P(x,x') = P(x) , \qquad \int \! dx \, P(x,x') = P(x') .
\end{equation}
Clearly $P(x,x')$ is not a proper probability density in general. However, when
interpreting the atomic fields as classical ones: $\hat{\phi}(x) \!\to\!
\phi(x)$, it becomes a proper classical joint probability density. In this
classical interpretation, we may now apply the Schwarz inequality that states
\begin{equation}
  \int \! dx \! \int \! dx' \, q(x) q(x') P(x,x') \leq \int \! dx \, q^2(x)
  P(x) \qquad \mbox{(classically)} .
\end{equation}
Applying this result to Eq.~(\ref{eq:C-def}) one obtains a classical inequality
for the contribution to the size of the atomic cloud:
$\sigma_q^2(t) \!\geq\! 0$. In consequence, via Eq.~(\ref{eq:dx}) one arrives
at the classical inequality for the size of the atomic cloud: $\Delta x_{\rm
a}(t) \!\geq\! \Delta X_{\rm s}$.

A violation of the Schwarz inequality would indicate true quantum
correlations between atoms as opposed to classical ones, since
then the second expectation value in Eq.~(\ref{eq:C-def}) cannot be
described by integration over a proper probability density. The
condition for this case is then
\begin{equation}
  \label{eq:cs-cond}
  \Delta x_{\rm a}(t) < \Delta X_{\rm s} \qquad \mbox{(Schwarz violation)} .
\end{equation}
Indeed, since Eq.~(\ref{eq:C-def}) contains atom-atom correlations in the
second term on the rhs, one may interpret a violation of the Cauchy--Schwarz
inequality, as formulated in Eq.~(\ref{eq:cs-cond}), as a test for true
quantum correlations between atomic pairs. Note, that the concept of defining
quantumness here is the same as in the case of photon antibunching in the
context of the optical Hanbury Brown--Twiss experiment.

A Schwarz violation at time $t$ can be easily shown to be equivalent to a
violation of the (classical) inequality
\begin{equation}
  \label{eq:example}
  \Delta q(t) \geq \sqrt{\langle [\Delta \hat{Q}(t)]^2 \rangle} ,
\end{equation}
where $\Delta q$ and $\Delta Q$ are the rms spread of the "single-atom"
quadrature, corresponding to the atomic density, and the rms spread of the cm
quadrature, respectively. The nature of such a violation can be
understood by considering the special case where the Schwarz violation occurs at
a time where the quadrature~(\ref{eq:q}) reduces to the atomic position, i.e.
$q(t) \!\to\! x$. The relation~(\ref{eq:example}) then states, that classically
the size of the atomic cloud $\Delta x$ is always equal or larger than the rms
spread of the cm of the cloud. In other words, the cm of the object is well
localized within the spatial extension of the cloud. A violation of this
classical inequality would then correspond to cases where the cm coordinate
reveals a rms spread larger than the cloud's size. In the extreme case, the
atomic cloud can then be seen as an almost pointlike object, its internal
distribution not being resolved, whose (cm) coordinate fluctuates.

It is thus the transition from an atomic cloud with well localized cm to a
quasi pointlike object with large fluctuation of its coordinate, that
corresponds to a transition of classical to quantum atom-atom correlations in
the cloud. Clearly for the general case, i.e. a Schwarz violation at an arbitary
time withing the half period $\pi/\omega$, the above interpretation
correspondingly holds for a specific quadrature $q(t)$ instead of position.

\begin{figure}
  \epsfig{width=12cm,file=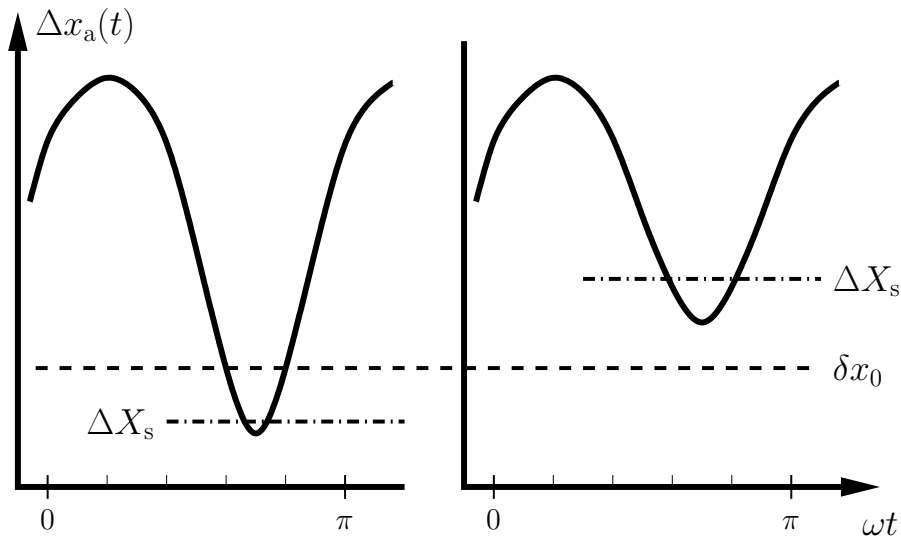}
  \caption{Observable size of the cloud versus time: For the parameter range
    given by Eq.~(\ref{eq:ranges}) (left part) the Schwarz boundary is below the
    one for QS whereas for values of $\eta$ outside this range (right part)
    the reverse is true.}
  \label{fig:sq-cs}
\end{figure}

In overall thus two levels of quantum or -- if one wishes to use this term --
non-classical behavior can be distinguished by the amplitude of the
oscillation of the atomic-cloud size. These two however, single-atom QS and
Schwarz violation, do not form a unique hierarchy: For different parameters
$\eta$ of the feedback mechanism the two
boundaries [cf.~Eqs~(\ref{eq:squeeze-cond}) and (\ref{eq:cs-cond})], appear in
different orders.  The order depends on whether $\Delta X_{\rm s} \!\leq\!
\delta x_0$ or not. The parameter range for this condition is obtained as
\begin{equation}
  \label{eq:ranges}
  \Delta X_{\rm s} \leq \delta x_0 \quad \mbox{for} \quad N \!-\! \sqrt{N^2
  \!-\! 1} \leq \eta \leq N \!+\! \sqrt{N^2 \!-\! 1} .
\end{equation}
In the case $N\!\to\! \infty$ this range includes all possible values of $\eta$
and thus for a truly macroscopic system, before a Schwarz violation
can be observed always first single-atom QS appears. This case is depicted
in the left part of Fig.~\ref{fig:sq-cs}.

However, for a finite system values for $\eta$ can be found outside the
range~(\ref{eq:ranges}) which then reveals an exchange of the order of
single-atom QS and Schwarz violation. In this case quantum correlations
between atoms
can be observed without the presence of single-atom QS, cf. right part of
Fig.~\ref{fig:sq-cs}. The corresponding parameter range is given by
\begin{equation}
  \label{eq:ranges2}
  \eta < N \!-\! \sqrt{N^2 \!-\! 1} \quad \mbox{or} \quad \eta > N \!+\!
  \sqrt{N^2 \!-\! 1} .
\end{equation}
In this range of $\eta$ apparently the quantumness of correlations can be
detected at much higher noise levels than the single-atom SQL
$\delta x_0$, which may render its experimental observation substantially
more feasible. Moreover, the two ranges in Eq.~(\ref{eq:ranges2}) correspond to
weak and strong feedback localization, respectively, and thus may allow
for a suitable combination of values for the shift rate $\zeta$
and the rms time-integrated measurement resolution $\sigma$ in an experiment.
Last but not least this scheme may be applied even for the extreme case of only
two indistinguished bosonic atoms, for which the range of possible values of
$\eta$, according to the case~(\ref{eq:ranges2}), becomes even more broad.

The criterium for quantumness of correlations can be tested directly from
measurements of the size of the atomic cloud $\Delta x(t)$ over half a period of
the trap oscillation, after the system has reached its asymptotic behavior
within a delay time of the order of $1/\zeta$. Experimentally a sufficiently
large number of sequences of feedback evolutions of varying time duration and
final cloud-size measurements have to be performed. Each sequence starts with
the identically prepared initial quantum state, whose correlations are to be
detected. Thus the final cloud-size measurements can be arbitrarily destructive
and can be performed for example by density-profile measurements.
These may be implemented by absorption imaging~\cite{absorb}, dispersive light
scattering~\cite{dispersive-scattering}, or possibly phase-modulation
spectroscopy~\cite{phase-spectroscopy}. Thus together with possible experimental
techniques to generate the required feedback loop~\cite{raithel,rempe}, an
experimental implementation of the presented scheme and a test for quantum
atom-atom correlations in bosonic gases seem to be feasible.

In summary we have shown that the quantumness of atom-atom correlations in a
trapped bosonic gas can be made observable as size oscillations of the atomic
cloud via feedback. For weak and strong feedback localization a Schwarz
violation for atom-atom correlations of a gas with finite atom number can be
observed in the absence of single-atom QS at correspondingly higher noise levels
than the SQL. Together with the feasibility of implementing this scheme with
present experimental techniques, this may allow for detecting the quantumness of
atom-atom correlations in a bosonic gas.

\acknowledgments
S.W. acknowledges support from the FONDECYT projects no. 1051072 and no.
7050184.

\end{document}